\documentclass[fleqn,usenatbib]{mnras}

\usepackage{newtxtext,newtxmath}

\usepackage{float}

\usepackage[T1]{fontenc}

\DeclareRobustCommand{\VAN}[3]{#2}
\let\VANthebibliography\thebibliography
\def\thebibliography{\DeclareRobustCommand{\VAN}[3]{##3}\VANthebibliography}

\usepackage{graphicx}	
\usepackage{amsmath}	
\usepackage{subfigure}
\usepackage{multirow}
\usepackage{multicol}
\usepackage{arydshln}
\usepackage{cite}
\usepackage{CJKutf8}

\usepackage[utf8]{inputenc}

\def\be{\begin{equation}}
\def\ee{\end{equation}}

\title[Polarisation features features of FRBs]{Polarisation of magnetospheric curvature radiation in repeating fast radio bursts}

\author[Wei-Yang Wang et al.]{Wei-Yang Wang (\begin{CJK*}{UTF8}{gbsn}王维扬\end{CJK*})$^{1,2}$\thanks{E-mail: wywang\_astroph@pku.edu.cn},
Jin-Chen Jiang (\begin{CJK*}{UTF8}{gbsn}姜金辰\end{CJK*})$^{3}$,
Kejia Lee (\begin{CJK*}{UTF8}{gbsn}李柯伽\end{CJK*})$^{2,3}$,
Renxin Xu (\begin{CJK*}{UTF8}{gbsn}徐仁新\end{CJK*})$^{1,2}$\thanks{E-mail: r.x.xu@pku.edu.cn},
\newauthor
Bing Zhang (\begin{CJK*}{UTF8}{gbsn}张冰\end{CJK*})$^{4,5}$\\
$^{1}$School of Physics and State Key Laboratory of Nuclear Physics and Technology, Peking University, Beijing 100871, People’s Republic of China\\
$^{2}$Kavli Institute for Astronomy and Astrophysics, Peking University, Beijing 100871, People’s Republic of China\\
$^{3}$National Astronomical Observatories, Chinese Academy of Sciences, 20A Datun Road, Chaoyang District, Beijing 100101, People’s Republic of China\\
$^{4}$Nevada Center for Astrophysics, University of Nevada, Las Vegas, NV 89154, USA\\
$^{5}$Department of Physics and Astronomy, University of Nevada, Las Vegas, NV 89154, USA\\
}

\date{Accepted XXX. Received YYY; in original form ZZZ}

\pubyear{2022}

\begin{document}
\label{firstpage}
\pagerange{\pageref{firstpage}--\pageref{lastpage}}
\maketitle

\begin{abstract}
Fast radio busts (FRBs) can exhibit a wide variety of polarisation properties, not only between sources but also from burst to burst for a same one.
In this work, we revisit the polarisation characters of coherent curvature radiation from a bulk of charged bunches in the magnetosphere of a highly magnetized neutron star.
FRBs have been observed to have a variety of polarisation features, such as high levels of circular polarisation or a sign change of circular polarisation.
High linear polarisation would appear when the line of sight is inside the emission beam (the on-beam geometry), whereas high circular polarisation would be present when it is outside (the off-beam geometry).
By considering two scenarios of the ``bulk shapes'' (thick vs. thin), we apply the model to explain the polarisation features of four repeating FRBs (FRB 20121102A, FRB 180916B, FRB 20190520B and FRB 20201124A). Most bursts are dominated by linear polarisation and negligible events have sign changes in circular polarisation, suggesting that such FRBs are most likely to be emitted by the ``thin'' bulks with large opening angles.
The higher probability of ``thin'' bulks could be meaningful for understanding repeating FRB central engine, i.e., the sparking dynamics to produce different bulks of energetic bunches on a neutron star surface.

\end{abstract}




\section{Introduction}\label{sec1}
Fast radio bursts (FRBs) are millisecond-duration radio flashes with extremely high bright temperatures (\citealt{Lorimer07,Thornton13}, also see \citealt{Cordes19,Petroff19} for reviews).
At present, hundreds of FRB sources have been discovered\footnote{Sources are catalogued on the Transient Name Server, https://www.wis-tns.org/.}, and a small proportion of them are repeaters.
Even though the observed sample is growing  and many mechanisms been proposed to explain the diverse observed properties (see \citealt{Platts19,Zhang20} for reviews), the underlying physical origin(s) of these bursts still remains an open question.

Polarisation measurements are tools to shed light on the possible radiation mechanisms of FRBs.
Most FRBs have linear polarisation (LP) fractions from dozens of percent up to 100\% \citep{Masui15,Michilli18,Luo20}, but significant circular polarisation (CP) has also been discovered in some bursts \citep{Masui15,Day20,Xu21,Kumar22,Jiang22}.
For some FRBs, the polarisation position angle (PA) for the LP remains constant across each burst \citep{Michilli18,Nimmo21}.
However, in some other FRBs, variable PAs across each burst have be observed, and the swing patterns are quite diverse among bursts \citep{Cho20,Luo20,Jiang22}.
These properties are reminiscent of the polarisation properties seen from magnetars (both transient radio bursts and normal pulsations, e.g., \citealt{Camilo16,Kirsten21}), which are a group of highly magnetized neutron stars. Notably, a mega-Jansky FRB-like burst was discovered from a known Galactic magnetar SGR J1935+2154 \citep{Bochenek20,CHIME20}.
Motivated by these observations, magnetars have emerged as the most likely origin for at least some repeating FRBs \citep{Beloborodov20,Ioka20,Lu20,Wadiasingh20,Yang21,Yuan22}.

There have been several mechanisms proposed to interpret these polarisation properties, of particular interest here is the CP characteristics.
Within the framework of magnetospheric curvature radiation model, we proposed that emission would have significant CP fractions if the line of sight (LOS) is not inside the emission beam \citep{Wang22a,Wang22}.
Other intrinsic mechanisms, e.g. inverse Compton scattering \citep{Zhang22}, may also create circularly polarized emission by adding many scattered linearly polarized waves \citep{Xu00}, however it is difficult for this model to reproduce observed CP fractions of tens of percent.
Models invoking propagation effects, e.g., multi-path processes \citep{Beniamini22} or the polarisation dependent radiative transfer mechanisms (e.g., Faraday conversion, see \citealt{Gruzinov19,Vedantham19,Kumar22b}) could also produce CP.
Observationally, polarisation profile oscillation with wavelengths is expected for the latter models. In any case, the outcoming wave through the plasma medium requires a highly circularly polarized incoming wave to interpret an FRB with high CP fractions.

In this paper, we mainly consider the intrinsic CP model, which invokes coherent curvature radiation by charged bunches.
Curvature radiation from charged bunches has been proposed to account for coherent radio emission of both pulsars (e.g., \citealt{RS75,Sturrock75,Elsaesser76,Cheng77,Melikidze00,Gil04,Gangadhara21}) and FRBs (e.g., \citealt{Katz14,Kumar17,Yang18,Ghisellini18,Katz18,Lu18,Wang20,Cooper21,Wang22}).
The model is developed from a work~\citep{Wang22} by deriving different geometric conditions of the emitting bulk.
We attempt to demonstrate the variety of polarisation properties of repeating FRBs that can be reproduced from this model.
Other intrinsic models invoking synchrotron maser coherent mechanism in a magnetar-wind-driven
external shock can interpret $\sim100\%$ LP but it is unclear how CP may be created in these models (e.g., \citealt{Metzger19}).
The paper is organized as follows.
We discuss polarisation and temporal properties in Section \ref{sec2}.
A comparison with observations and some implications are demonstrated in Section \ref{sec3}.
The results are discussed and summarized in Section \ref{sec4}.
The convention $Q_x=Q/10^x$ in cgs units is used throughout the paper.

\section{polarisation properties}\label{sec2}
\subsection{Coherent curvature radiation by bunches}

\begin{figure}
\centering
\includegraphics[width=0.48\textwidth]{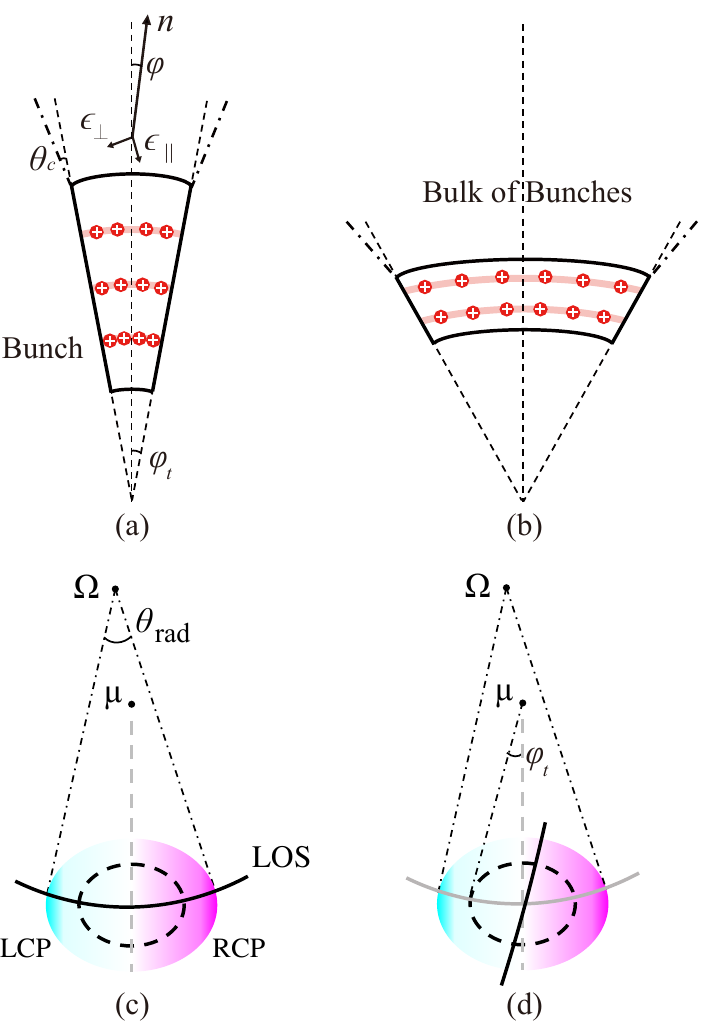}
\caption{The upper two diagrams denote schematic diagrams of a bulk of bunches: (a) a ``thick'' bulk for $\theta_{\text {rad }}/\Omega \ll t_{\text {int }}$; (b) a ``thin'' bulk for $\theta_{\text {rad }}/\Omega \gg t_{\text {int }}$.
Dashed-dotted lines denote conal radiation regions.
The half opening angle of the bulk is $\varphi_t$.
The light
solid red lines show the slice (bunch) in which charges emit at roughly the same phase. Bunches are assumed to carry net positive charges. The lower two diagrams denote the
schematic diagram of the emission beam observed in frame of a moving bulk: (c) a ``thick'' bulk; (d) a ``thin'' bulk.
Different colors denote the degree of LCP (cyan) and RCP (magenta).
The black solid lines are LOSs.
For the ``thin'' bulk case, bunched particles have traveled through the emission region faster than the LOS sweeping the conal radiation region, so that the LOS sweeps from the left bottom to the right top along the solid line, rather than the grey line in panel (d).
The grey dashed lines is the bulk central axis where $V=0$.
}
\label{fig1}
\end{figure}

Within the coherent curvature radiation model of FRBs, FRBs are triggered by a sudden and violent ``sparking'' process in contrast to a continuous process  required to power a normal pulsar \citep{RS75}.
These sparking particles can form charged bunches, such that the total observed radiation is coherently enhanced significantly when the size of the charged bunch is smaller than the half-wavelength in the observer frame.
The bunch formation mechanisms have been explored extensively in the pulsar context within the neutron star magnetosphere, which generally invoke a two-stream instability (e.g., \citealt{Usov87,Asseo98,Melikidze00,Ben21,Kumar22a}). The mechanism for FRBs may be similar but the bunch needs to have a much larger number of net charges so that FRB emissions are much brighter.

Since particle momentum perpendicular to the field line drops to zero rapidly, the particle trajectories essentially track with the magnetic field lines.
Charged particles, which may be produced by a sudden trigger in the inner gap, stream outwards along curved magnetic field lines and emit curvature photons.
An electric field $E_\|$ parallel to the magnetic field line may exist in the charge starvation region to continuously inject energy to the bunches to maintain the observed luminosity
for the typical FRB duration \citep{Kumar17}.
The balance between the radiation power and $E_\|$ can be established quickly \citep{Wang19}, therefore, a constant Lorenz factor distributed in a charged bunch is considered in our following calculation.

The observed emission intensity cannot be simply demonstrated by the summation of the curvature radiation amplitude of individual particles, because FRB emissions are significantly coherent.
Basically, curvature radiation from a single charge can be described by two orthogonal polarized components, i.e., $A_\|$ and $A_{\perp}$, in which $A_\|$ is earlier than $A_\perp$ by $\pi/2$ in phase.
The unit vector $\boldsymbol{\epsilon}_\|$ is pointing to the direction of the instantaneous curvature radius of the field line, 
and $\boldsymbol{\epsilon}_{\perp}=\boldsymbol{n} \times \boldsymbol{\epsilon}_{\|}$ is defined, where $\boldsymbol{n}$ denotes the unit vector of the LOS.
For a single charge with identifiers $i$, $j$, $k$, we define
\begin{equation}
\xi=\frac{\omega \rho}{3 c}\left(\frac{1}{\gamma^{2}}+\varphi_{k}^{2}+\chi_{i j}^{2}\right)^{3 / 2},
\end{equation}
where $\omega$ is the angular frequency, $\gamma$ is the Lorentz factor, $\rho$ is the curvature radius, $\chi_{ij}$ is the angle between the considered trajectory and the trajectory at $t = 0$, $\varphi_k$ is the angle between the LOS and the trajectory plane.
The critical angular frequency of curvature radiation is defined as $\omega_{c}=3 c \gamma^{3} /(2 \rho)$.
The amplitudes for one charged particle are given by
\be
\begin{aligned}
A_{\|, i j k} \simeq & \frac{i 2}{\sqrt{3}} \frac{\rho}{c}\left(\frac{1}{\gamma^{2}}+\varphi_{k}^{2}+\chi_{i j}^{2}\right) K_{\frac{2}{3}}(\xi) \\
&+\frac{2}{\sqrt{3}} \frac{\rho}{c} \chi_{i j}\left(\frac{1}{\gamma^{2}}+\varphi_{k}^{2}+\chi_{i j}^{2}\right)^{1 / 2} K_{\frac{1}{3}}(\xi), \\
A_{\perp, i j k} \simeq & \frac{2}{\sqrt{3}} \frac{\rho}{c} \varphi_{k}\left(\frac{1}{\gamma^{2}}+\varphi_{k}^{2}+\chi_{i j}^{2}\right)^{1 / 2} K_{\frac{1}{3}}(\xi),
\end{aligned}
\ee
where $K_\nu(\xi)$ is the modified Bessel function \citep{Jackson98}.

We introduce the concept of ``bulk of bunches'' to closely discuss their dynamical and radiation properties, as shown in Figure \ref{fig1}.
A parallel electric field may form and propagate outward like a travelling wave along the unperturbed magnetic field, so that the charged particles are accelerated to move in the same direction \citep{Kumar22}.
Charges in a bunch are suggested to move along nearly identical orbits, therefore they act like a single macro charge.
The emission from a charge is coherently added within one bunch, i.e., a power is proportional to $N_e^2$, where $N_e$ is the number of net charges in one bunch.
Within a bulk, on the other hand, there could be $N_{lb}$ bunches contributing to the observed instantaneous radiation, with the emissions from them added incoherently.
The total energy radiated per unit solid angle per unit frequency interval can be written as
\begin{equation}
\frac{d^{2} W}{d \omega d \Omega}=N_{l b} \frac{e^{2} \omega^{2}}{4 \pi^{2} c}\left|\sum_{i}^{N_{l}} \sum_{j}^{N_{\theta}} \sum_{k}^{N_{\phi}}-\boldsymbol{\epsilon}_{\|} A_{\|, i j k}+\boldsymbol{\epsilon}_{\perp} A_{\perp, i j k}\right|^{2},
\end{equation}
where $(i,\,j,\,k)$ are three subscripts to identify a particle, and the number of net charges in one bunch is $N_e=N_lN_{\theta}N_{\phi}$ \citep{Wang22}.

The spectra can evolve as bunches move and the line of sight sweeps, which are generally characterized by multisegmented broken power laws \citep{Wang22}.
Drifting pattern is a natural consequence of magnetospheric curvature radiation \citep{Wang19,Wang20}.
The amplitude of $A_{\perp}$ has been investigated by invoking off-beam LOS, which may lead to CP \citep{Wang22}.
In this paper, we focus on polarisation features of curvature radiation by deriving different ``shapes'' of the emitting bulk.

\subsection{Burst duration}

Let us consider that the LOS is inside the emission beam of the bunches.
The intrinsic duration, $t_{\rm int}$, of an FRB observed in the co-rotation frame is determined by the number of bunches that continuously sweep across the LOS.
Thus, the total number of persistent bunches traveling through the emitting region during the FRB emission can be estimated as $N_B\simeq 2\times10^6\nu_9t_{\rm int,-3}$~\citep{Wang22}, where $t_{\rm int}$ is the burst width and $\nu=\omega/(2\pi)$.

Even if the radial size is limited by the half-wavelength, the transverse size can be much larger.
Emission from such ultra-relativistic particles is mainly confined in a conal region.
The angle of the emission cone for a bunch is defined as $\theta_b=\varphi_t+\theta_c$, where
\begin{equation}
\theta_{c}(\omega) \simeq\left\{\begin{array}{cl}
\frac{1}{\gamma}\left(\frac{2 \omega_{c}}{\omega}\right)^{1 / 3}=\left(\frac{3 c}{\omega \rho}\right)^{1 / 3}, & \omega \ll \omega_{c} \\
\frac{1}{\gamma}\left(\frac{2 \omega_{c}}{3 \omega}\right)^{1 / 2}, & \omega \gg \omega_{c}
\end{array}.\right.
\end{equation}
The angle can be estimated as $\theta_b\simeq\varphi_t+1/\gamma$ at $\omega=\omega_c$.
Note that we define $\varphi < \theta_b$ as on-beam and $\varphi > \theta_b$ as off-beam.
This definition is more general than treating $\theta_c$ as the angle of the emission cone in \citet{Wang22}, by considering the transverse bunch size.

Emitting bunches essentially corotate with the magnetosphere.
The observed duration of an FRB in an observer frame reads
\be
w \simeq (1+z) \min \left( t_{\text {int }}, \theta_{\text {rad }}/\Omega \right),
\label{eq_width}
\ee
where $\theta_{\text {rad }}$ is the radiation beaming angle shown as panel (c) of Figure \ref{fig1}, $\Omega$ is the angular frequency of the neutron star, and $z$ is the redshift \citep{Yang19}.
Spherical coordinates $(r,\,\theta,\,\varphi)$ with respect to the magnetic axis and $(r,\,\Theta,\,\Phi)$ with respect to the spin axis are used.
If the LOS can sweep the beam center, the radiation beaming angle can be written as
\be
\theta_{\text{rad}}=2(\Phi_t+\theta_c),
\label{eq_jet}
\ee
where
\be
\sin\Phi_t=\frac{\sin\varphi_t\sin\theta}{\sin\zeta},
\ee
in which $\varphi_t$ is the half opening angle, and $\zeta$ is the angle between the LOS and the spin axis.

We consider two possible scenarios of the bulk of bunches.
According to Equation (\ref{eq_width}), one can define $\theta_{\text {rad }}/\Omega \ll t_{\text {int }}$ as a ``thick'' bulk 
and $t_{\text {int }}\ll \theta_{\text {rad }}/\Omega$ as a ``thin'' bulk, as shown in Figure \ref{fig1}.
The ``thickness'' here is determined by radial and transverse observing time rather than the true spatial size for the two dimensions.
So the ``thick'' and ``thin'' cases here are of the visual effects from an observer, not representing the intrinsic geometry of the bulk itself.
If the angle $\theta_{\rm rad}$ is almost constant, a ``thick'' bulk would be observed for a rapidly spinning object, while a ``thin'' bulk would be observed for a slowly spinning one.

\subsection{Polarisation Profile}

\begin{figure*}
\centering
\includegraphics[width=0.96\textwidth]{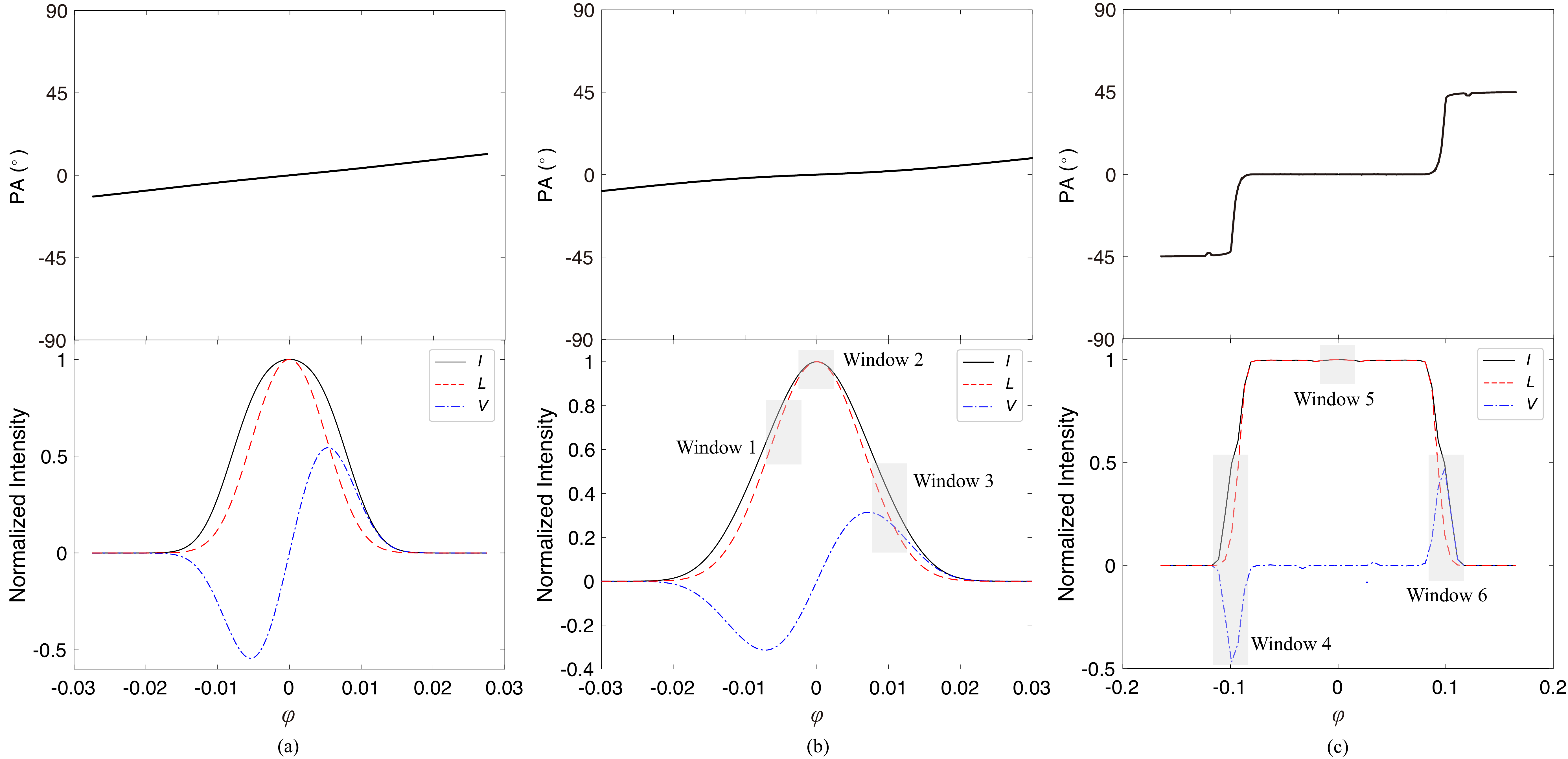}
\caption{Simulated polarisation profiles for  ``thick'' bulks: (a) $\varphi_t=0.1/\gamma$; (b) $\varphi_t=1/\gamma$; (c) $\varphi_t=10/\gamma$.
Top panels: The PA envelope across the burst in black solid line.
Bottom panels: The Stokes parameters $I$ and $V$ are plotted in black solid and blue dotted-dashed curves. They are normalized to the value of $I$ at $\varphi=0$.
The linearly polarized component $L$ is plotted in red dashed curves.
The parameters are adopted as $\gamma=100$ and $\omega=\omega_c$.
The grey regions show six observational windows as examples.
The width of each observational window is $2\sigma_w$.}
\label{fig2}
\end{figure*}

\begin{figure}
\centering
\includegraphics[width=0.48\textwidth]{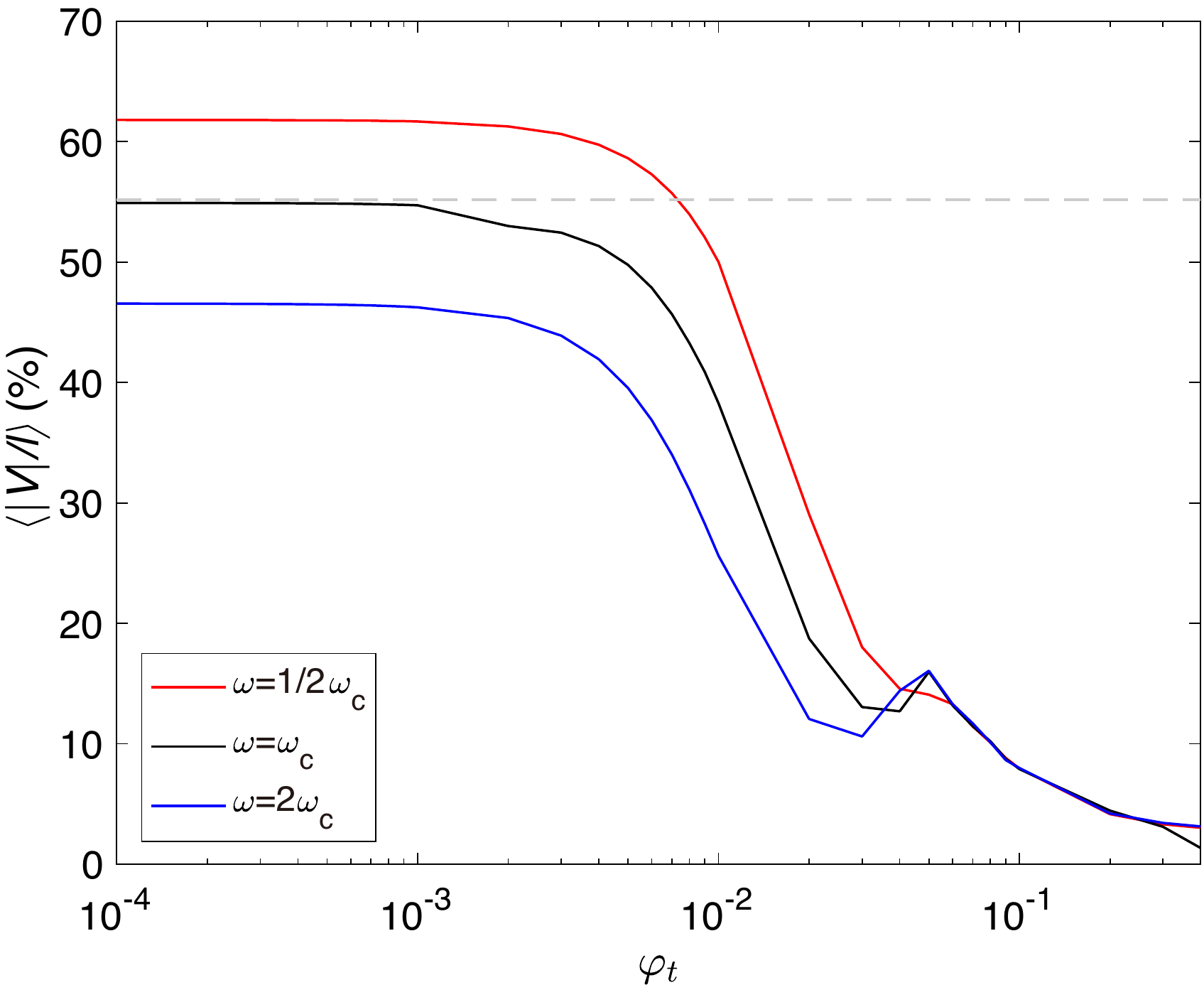}
\caption{The average CP fraction as a function of opening angle with different frequencies: $\omega=1/2 \omega_c$ (red solid curve), $\omega= \omega_c$ (black solid curve), $\omega=2 \omega_c$ (blue solid curve). The grey dashed line denotes the average CP fraction of a single charge for $\omega=\omega_c$.}
\label{fig6}
\end{figure}

Polarisation properties can reveal the information about the particle trajectories and the magnetic configuration. The evolution patterns of the Stokes parameters as the LOS sweeps the emission beam, i.e., the polarisation profiles, are derived to study the polarisation properties of bunched curvature radiation.
The Stokes parameters can be calculated as
\be
\begin{aligned}
I &=\mu\left(A_{\|} A_{\|}^{*}+A_{\perp} A_{\perp}^{*}\right) \\
Q &=\mu\left(A_{\|} A_{\|}^{*}-A_{\perp} A_{\perp}^{*}\right) \\
U &=\mu\left(A_{\|} A_{\perp}^{*}+A_{\perp} A_{\|}^{*}\right) \\
V &=-i \mu\left(A_{\|} A_{\perp}^{*}-A_{\perp} A_{\|}^{*}\right)
\end{aligned}
\ee
where $\mu=\omega^{2} e^{2} /\left(4 \pi^{2} \mathcal{R}^{2} c T\right)$ is the proportionality factor. The factor is such chosen that $I$ is the flux density averaged over a timescale $T$, and $\mathcal{R}$ is the distance from the emitting source to the observer.
The corresponding linearly polarized component and the PA read
\be
\begin{aligned}
L & =\sqrt{Q^2+U^2},\\
\psi & =\frac{1}{2} \tan ^{-1}\left(\frac{U}{Q}\right).
\end{aligned}
\ee

We assume that the curvature radius is a constant in the bulk and $\chi'=0.001$.
In the emitting bulk, 
There are $N_b$ bunches that can contribute to instantaneous radiation, whose electric fields are added incoherently.
Therefore, the dimensionless parameter denoting the enhancement factor due to coherence is $F_{\omega}\simeq N_e^2N_b$ \citep{Yang18,Wang22}.
Assuming that charges are normally distributed in $\chi'$ and $\varphi'$, the total amplitudes of the bulk are given by
\begin{equation}
\begin{aligned}
A_{\|} &\simeq \frac{1}{\sqrt{3}} \frac{\rho}{c} \frac{N_eN_b^{1/2}}{\varphi_t}  \int_{\varphi_d}^{\varphi_u}\left[i \chi^{\prime 2} K_{\frac{2}{3}}(\xi)+\chi^{\prime}\left|\chi^{\prime}\right| K_{\frac{1}{3}}(\xi)\right] \\
&\times\cos \varphi^{\prime} d \varphi^{\prime}, \\
A_{\perp} &\simeq \frac{1}{\sqrt{3}} \frac{\rho}{c} \frac{N_eN_b^{1/2}}{\varphi_t}  \int_{\varphi_d}^{\varphi_u}\left|\chi^{\prime}\right| K_{\frac{1}{3}}(\xi) \varphi^{\prime} \cos \varphi^{\prime} d \varphi^{\prime},
\end{aligned}
\label{eq:AA}
\end{equation}
where $\varphi_u=\varphi_t+\varphi$ and $\varphi_d=-\varphi_t+\varphi$.

In general, curvature radiation by bunches is 100\% polarized.
If charges are uniformly distributed in bunches, the emission would be 100\% linearly polarized when the LOS is parallel to the central axis.
We define $V<0$ as the left circular polarisation (LCP) and $V>0$ as the right circular polarisation (RCP), as shown in Figure \ref{fig1}.
The sign of $A_{\perp}$ would change when the LOS sweeps the central axis of the bulk, leading to sign change of $V$.
Significant CP can generate in the off-beam cases due to the nonaxisymmetric summation of $A_{\perp}$ \citep{Wang22a,Wang22}.

In general, highly circularly polarized waves can appear at off-beam cases, which tend to have lower fluxes than the on-beam cases under the same condition.
However, from Equation (\ref{eq:AA}), one can see that the emission amplitudes sensitively depend on $N_e$, $N_b$ and $\varphi_t$.
The sparking process can create charged particles with random numbers and bulk sizes, leading to large fluctuations of $N_e$, $N_b$ and $\varphi_t$.
Therefore, for the burst waves with a certain CP, the value of the observed flux may have a scattered distribution.

\subsubsection{Polarisation profile for a ``thick'' bulk}\label{sec2.3.1}
For a ``thick'' bulk, the time for persistent bunches to travel through the emitting region is much longer than that for the LOS to sweep the whole $\theta_{\rm rad}$ due to rotation, thus one can observe emission from the entire radiation beaming region, as shown in panel (c) of Figure \ref{fig1}.
We simulate the polarisation profile for a ``thick'' bulk in three opening angle cases ($\varphi_t=0.1/\gamma$, $\varphi_t=1/\gamma$, $\varphi_t=10/\gamma$) at $\omega=\omega_c$.
Since the flux drops to a small number rapidly when $\chi\gg1/\gamma$ and $\varphi\gg1/\gamma$, it is required that either $\chi\ll1/\gamma$ or $\varphi\ll1/\gamma$.
For simplicity, we assume $\chi=0.001$ here.
The Stokes parameters are considered in the spherical coordinates with respect to magnetic axis for a general discussion.

The simulated polarisation profiles are shown in Figure \ref{fig2}.
Emissions for all three cases retain high levels of LP as the LOS is inside the beam within an angle of $\theta_b$ (on-beam), and the CP fraction becomes significant when the LOS is off-beam.
The case with $\varphi<1/\gamma$ shares similar polarisation properties with that for $\varphi\sim1/\gamma$ (e.g., \citealt{Tong22}).
Waves are LCP at $\varphi<0$ but change to RCP when $\varphi>0$.
If $\varphi\gg1/\gamma$, there is a large phase space where the summation of $A_\perp$ cancels out, so that the emission has roughly 100\% LP when the LOS is inside the large beam angle.
A rapid polarisation conversion from LP to CP occurs at $|\varphi\simeq\theta_b|$ and emission becomes $\sim100\%$ CP for the off-beam case.

The average LP and CP fractions within the pulse width $\left\langle L/I\right\rangle$ and $\left\langle |V|/I\right\rangle$ are adopted to characterize the polarisation properties.
For $\omega=\omega_c$, the average CP fraction is $\left\langle |V|/I\right\rangle\approx38\%$ for $\varphi_t=1/\gamma$ and is smaller than $10\%$ when $\varphi\gg1/\gamma$.
The average CP fraction can reach 55\% when $\varphi_t\ll1/\gamma$, which is the same as $\left\langle |V|/I\right\rangle$ of a single charge within $-1/\gamma<\varphi<1/\gamma$ at $\omega=\omega_c$.
Charged bunches with $\varphi_t\lesssim1/\gamma$ can share similar polarisation properties as a single charge.
In general, the average CP fraction continuously decreases as the opening angle of the bulk increases.
We simulate the average CP fraction as a function of $\varphi_t$ in three frequency cases ($\omega=1/2 \omega_c$, $\omega= \omega_c$, $\omega=2 \omega_c$) as shown in Figure \ref{fig6}.
A burst tends to have a smaller $\left\langle V/I\right\rangle$ in higher frequencies because the radiation beaming angle decreases with frequency.
The average CP fraction tends to be the same regardless of the emission frequency if $\varphi_t\gg1/\gamma$.

\subsubsection{Polarisation profile for a ``thin'' bulk}

\begin{figure*}
\centering
\includegraphics[width=0.96\textwidth]{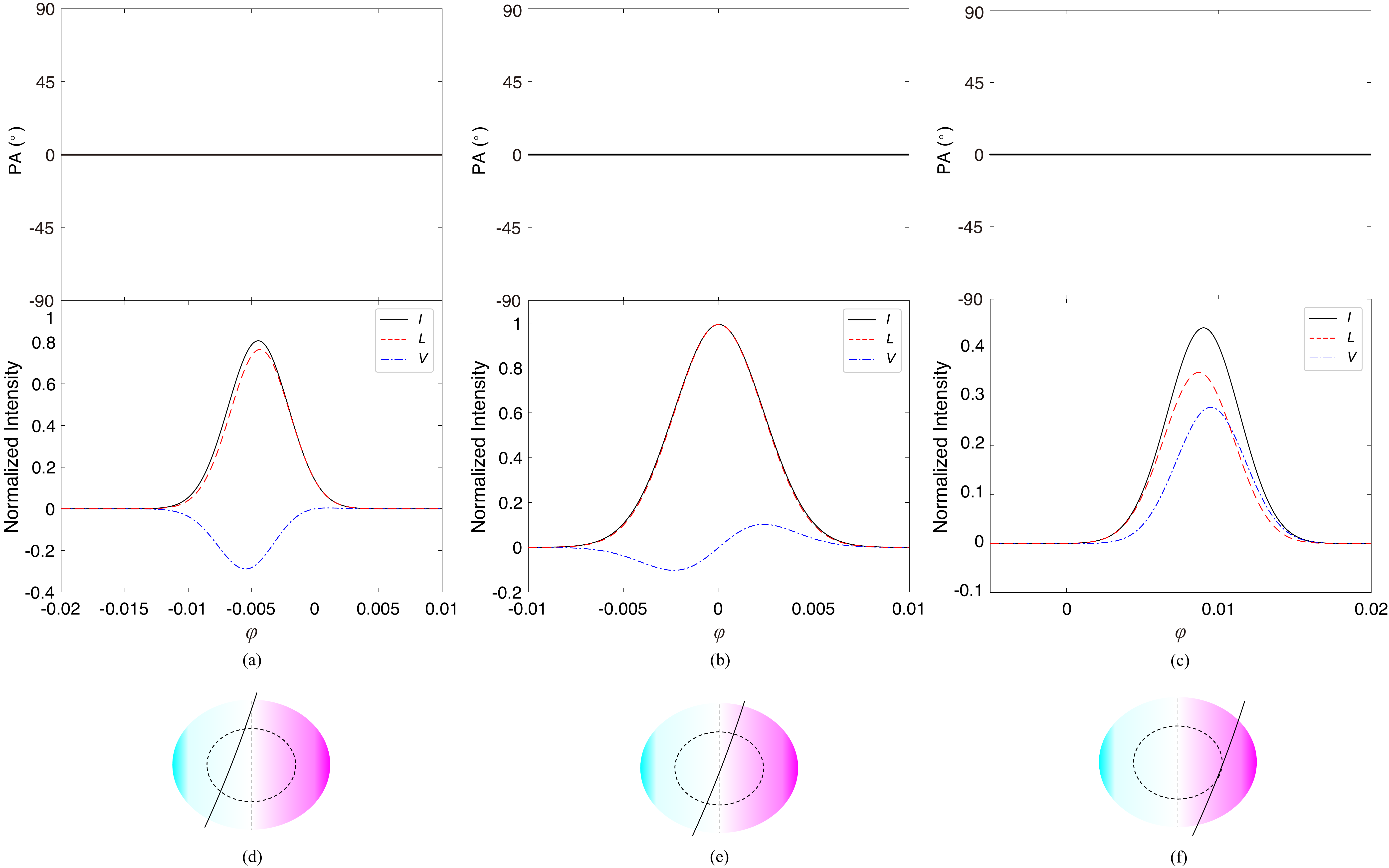}
\caption{Simulated polarisation profile for the ``thin'' bulks with $\varphi_t=0.01$: (a) $\varphi_p=-0.005$; (b) $\varphi_p=0$; (c) $\varphi_p=0.01$. Three observational windows from 1 to 3 are shown in the Figure \ref{fig2} panel (b).
Each schematic diagram of the beam is similar to panel (c) and (d) of Figure \ref{fig3}.
The PA across the burst envelope has a flat shape and is roughly equal to zero throughout.}
\label{fig3}
\end{figure*}

\begin{figure*}
\centering
\includegraphics[width=0.96\textwidth]{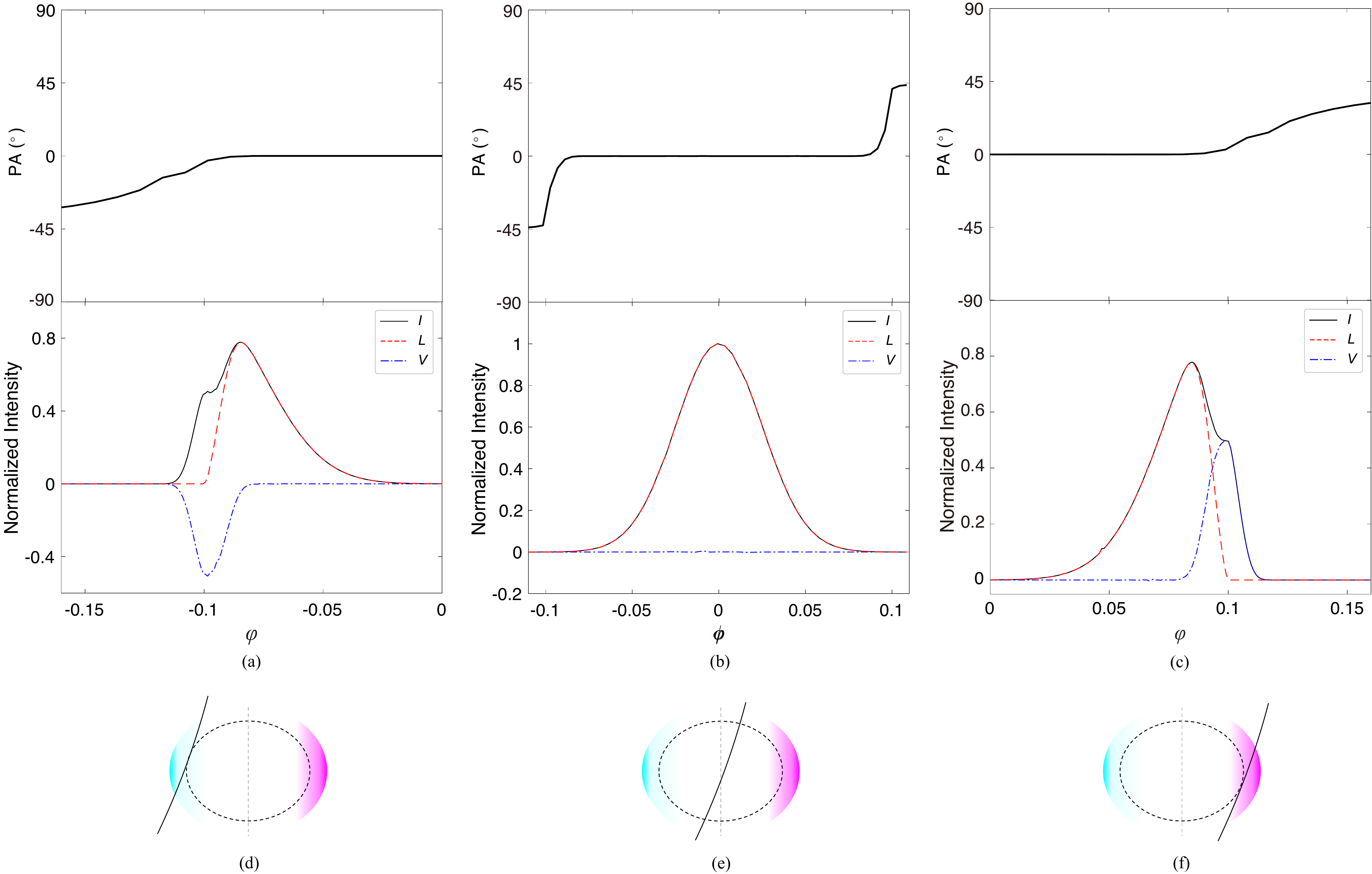}
\caption{The same as Figure \ref{fig3} (i.e. ``thin'' bulks), but for $\varphi_t=0.1$: (a) $\varphi_p=-0.1$; (b) $\varphi_p=0$; (c) $\varphi_p=0.1$. Three observational windows from 4 to 6 are shown in Figure \ref{fig2} panel (c).}
\label{fig4}
\end{figure*}

\begin{figure*}
\centering
\includegraphics[width=0.96\textwidth]{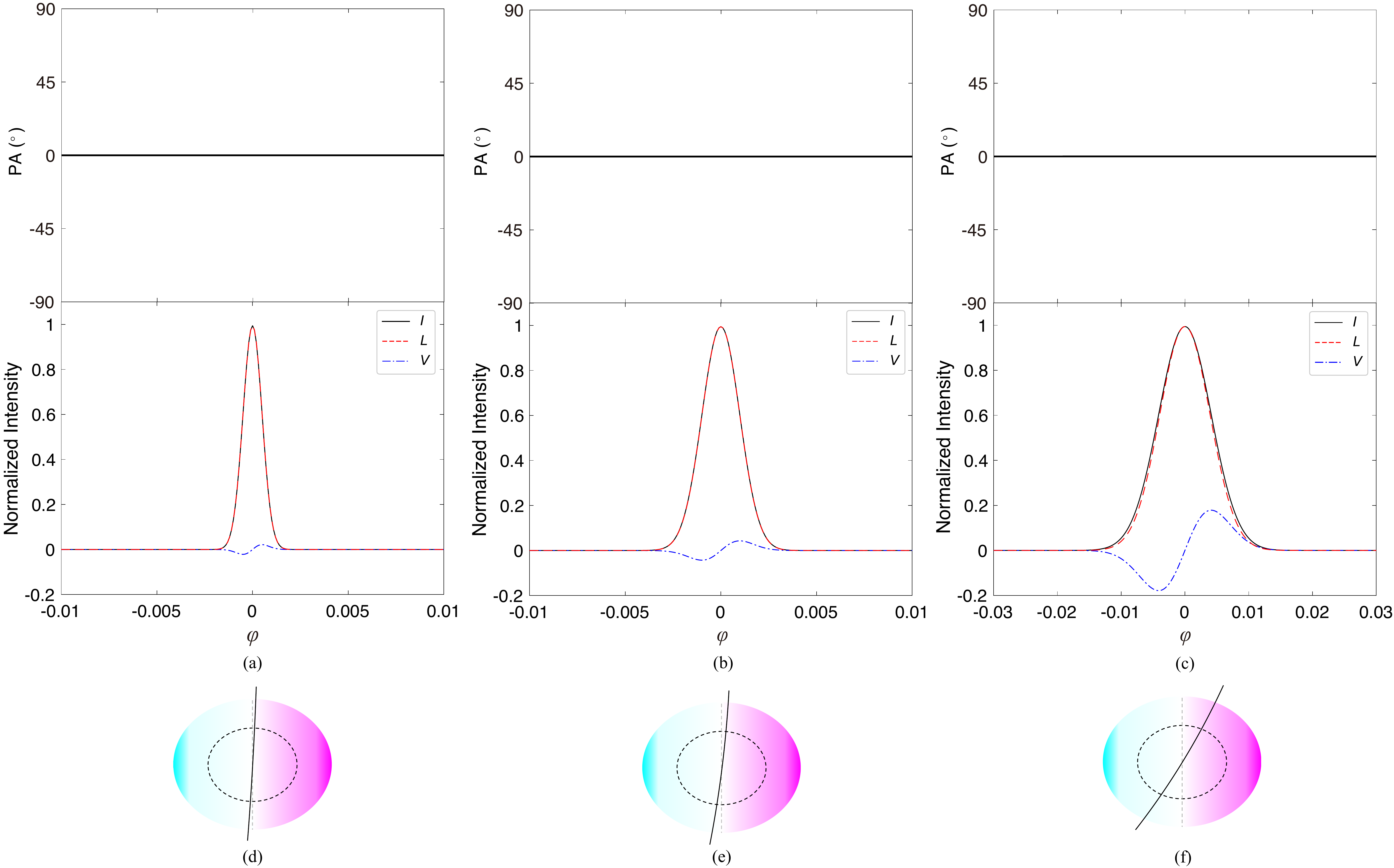}
\caption{The same as Figure \ref{fig3} (i.e. ``thin'' bulks), but for different widths of the observational window: (a) $\sigma_w=0.001$; (b) $\sigma_w=0.002$; (c) $\sigma_w=0.01$. The Gaussian peaks are adopted as $\varphi_p=0$.
Again the PA across the burst envelope is flat and is roughly equal to zero throughout.}
\label{fig5}
\end{figure*}

For a ``thin'' bulk, all the bunches have traveled through the emission region before the LOS sweeps the whole radiation beaming region, so that one may observe a significant part of the radiation cone.
The trajectories of LOS do not rotate around the spin axis in the frame of the moving bulk, as shown in the panel (d) of Figure \ref{fig1}.
Noticeably different polarisation properties from the ``thick'' bulk scenario will show up, since the width of observational window is narrower than $\theta_{\text{rad}}$.

The triggered photon-pair production cascade is a complex process.
We assume that the number density of pairs obeys a Gaussian function in terms of time and the LOS can sweep at the beam center.
Thus, the observed electric vectors are modulated by the number density when the bunches travel through the emission region.
The modulation function for the observed electric vectors can be written as
\begin{equation}
f(\varphi)=f_{0} \exp \left[-\left(\frac{\varphi-\varphi_{\mathrm{p}}}{\sigma_{w}}\right)^{2}\right],
\end{equation}
where $\varphi_p$ is the peak location of the Gaussian function, $f_{0}$ is the amplitude and $\sigma_{w}$ is the Gaussian width.
We fix $f_0=1$ in the following calculation.

Consider observational windows appearing at different phases within the radiation beaming region.
Adopting $\sigma_w=0.005$, we simulate the polarisation profile and PA across the burst envelope in three cases ($\varphi_p=-0.005$, $\varphi_p=0$ and $\varphi_p=0.01$), under the assumption of $\varphi_t=0.01$, $\gamma=100$ and $\omega=\omega_c$ as shown in Figure \ref{fig3}.
The width of observational window is $2\sigma_w$, as shown in panel (b) of Figure \ref{fig2}.
A wide variety of polarisation properties between observational widows may be exhibited.
The waves have a purely LCP in panel (a) and a purely RCP in panel (c).
The sign of CP will change when the symmetric axis of the bulk is inside the observational window.
The average CP fraction for panels (a), (b) and (c) of Figure \ref{fig3} are $32.2\%$, $13.5\%$ and $60.2\%$, respectively, but all of them have constant PAs across the burst envelope.
We assume that the observational window is located at one side of the bulk central axis.
The intensity of $A_{\|}$ becomes smaller and be close to $A_{\perp}$ as the window gets wider.
There may be highly circularly polarized waves ($\left\langle |V|/I\right\rangle>55\%$ at $\omega=\omega_c$) being seen if the LOS only sweeps part of $\theta_{\rm rad}$.

For a ``thin'' bulk but with $\varphi_t=0.1$, the simulation in three cases ($\varphi_p=-0.1$, $\varphi_p=0$ and $\varphi_p=0.1$) are shown in Figure \ref{fig4} and the observational windows are shown in panel (c) of Figure \ref{fig2}.
The average CP fraction for panels (a), (b) and (c) of Figure \ref{fig4} are $25.4\%$, $0\%$ and $25.4\%$, respectively.
The sign of CP does not change regardless of the width of the observational window.
If the center of the observational window is normally distributed within in the radiation beaming region, one would have a high probability of seeing a $\sim100\%$ linearly polarized burst.
The chance for detecting CP decreases as the opening angle becomes larger.

Consider that the Gaussian peak overlaps with the central axis of the bulk coincidently.
The width of the observational window can be different.
Adopting $\varphi_p=0$, we simulate the polarisation profile and PA across the burst envelope in three cases ($\sigma_w=0.001$, $\sigma_w=0.002$ and $\sigma_w=0.01$), under the assumption of $\varphi_t=0.01$, $\gamma=100$ and $\omega=\omega_c$ as shown in Figure \ref{fig5}.
The average CP fraction for panels (a), (b) and (c) of Figure \ref{fig5} are $2.8\%$, $5.6\%$ and $23.3\%$, respectively.
Within the pulse window, all three panels in Figure \ref{fig5} exhibit sign change of CP.
One can infer that the average CP fraction increases as the observational window gets wider when the Gaussian peak appears at the central axis of the bulk with $\varphi_t\lesssim1/\gamma$.

The polarisation profiles have the largest derivative of CP fraction with respect to $\varphi$ at $\varphi=0$.
This quantity can reach the largest value when it is a single charge:
\be
P\frac{\Delta(V/I)}{\Delta t}<\frac{d(V/I)}{d\varphi}\lesssim1.24\gamma,
\label{eq11}
\ee
where $P$ is the period of the neutron star.
From Equation (\ref{eq11}), one can obtain $P\lesssim1.24\gamma_2\Delta t_{-3}[\Delta (V/I)]^{-1}_{-1}\rm\,s$.
Note that this constraint is independent of whether the emitting bulk is ``thin'' or ``thick''.
In order to generate ``thin'' bulks, combining Equation (\ref{eq_width}) and (\ref{eq_jet}), one can derive the period of the neutron star $P>0.3\gamma_2w_{-3}$ s.
The neutron star may be a slowly rotating pulsar or a magnetar\footnote{
It could be extremely difficult to measure the spin period of these slowly rotating neutron stars manifested in the form of FRBs, compared with that of regular radio pulsars.
(1) The radiation window would be large, though for a slow rotator, due to emission at low altitude of a star with significant multi-pole magnetic fields, and a radio burst with  millisecond-duration may appear almost randomly in the window.
(2) Large timing irregularities could also result from an enhanced spindown caused by high radiation power in clean magnetosphere~\citep{Wang22a}, as well as quake-induced activity.
}.%

\section{Implications from observations}\label{sec3}

Bursts from most FRBs are polarized.
They exhibit noticeable differences between different sources and different bursts from the same source.
Some polarisation properties of the well-observed repeaters have been summarized in Table \ref{tab1}.
In general, the LP fraction of the bursts depends on frequency, which shows a trend of lower LP at lower frequencies \citep{Feng22}.
The total polarisation degree of the intrinsic model is 100\%.
We discuss the bursts with higher frequencies since depolarisation due to the multi-path effect is not significant.
However, $\sigma_{\rm RM}$ is different for each source, so there is no common cut-off frequency for all FRBs.
Among the four FRB sources (FRB 20121102A, FRB 20180916B, FRB 20190520B and FRB 20201124A) that have a large sample of polarisation measurements, FRB 20201124A keeps $\sim100\%$ total polarisation degree for $\nu>1 \rm\,GHz$, similar to FRB 20121102A at $\nu>3 \rm\,GHz$.
Other two sources were mostly observed at frequencies where depolarisation is significant.
We attempt to discuss some polarisation properties by considering the intrinsic radiation mechanism for the two sources.

The Stokes parameters for most of these bursts do not oscillate with wavelength.
The polarisation-dependent radiative transfer mechanism may work for some bursts but not all of them.

\begin{table*}
\begin{center}
\caption{Some polarisation properties of repeaters}
\begin{tabular}{cccccc}
\hline \hline
FRB Source & Band (GHz) & LP (\%) & CP (\%) & PA & Ref.\\
\hline
20121102A &  3--5 & $\sim100$ & 0 & Constant$^{\rm a}$ & \citet{Michilli18,Hilmarsson21} \\
 & 1--1.5 & $<20$ & $<15$ & - & \citet{Plavin22} \\
20180301A & 1-1.5 & 36--80 & $<10$ & Either constant or varying & \citet{Luo20}\\
20180916B & 0.3--1.7 & $\gtrsim80$ & $\lesssim15$ & Constant & \citet{Nimmo21,Sand21} \\
& 0.1--0.2 & 30--70 & 0 & Constant & \citet{Pleunis21} \\
20190303A & 0.4-0.8 & $\gtrsim20$ & - & Constant & \citet{Fonseca20}\\
20190417A & 1--1.5 & 52--86 & - & Constant & \citet{Feng22}\\
20190520B & 2.8--8 & 15--80 & $<15^{\rm b}$ & Constant & \citet{Anna22,Dai22,Niu22a}\\
20190604A & 0.4-0.8 & $\sim100$ & 0 & Constant & \citet{Fonseca20}\\
20201124A & 0.7--1.5 & 30--100 & 0--90 & Either constant or varying & \citet{Xu21,Kumar22,Jiang22}\\
\hline \hline
\end{tabular}
\label{tab1}
\end{center}
$^{\rm a}${Only one burst of FRB 20121102A has varying PA \citep{Hilmarsson21}.\\}
$^{\rm b}${There is one burst showing CP fraction of $42\pm7\%$ \citep{Anna22}.\\}
\end{table*}

\subsection{FRB 20201124A}

\begin{figure}
\centering
\includegraphics[width=0.48\textwidth]{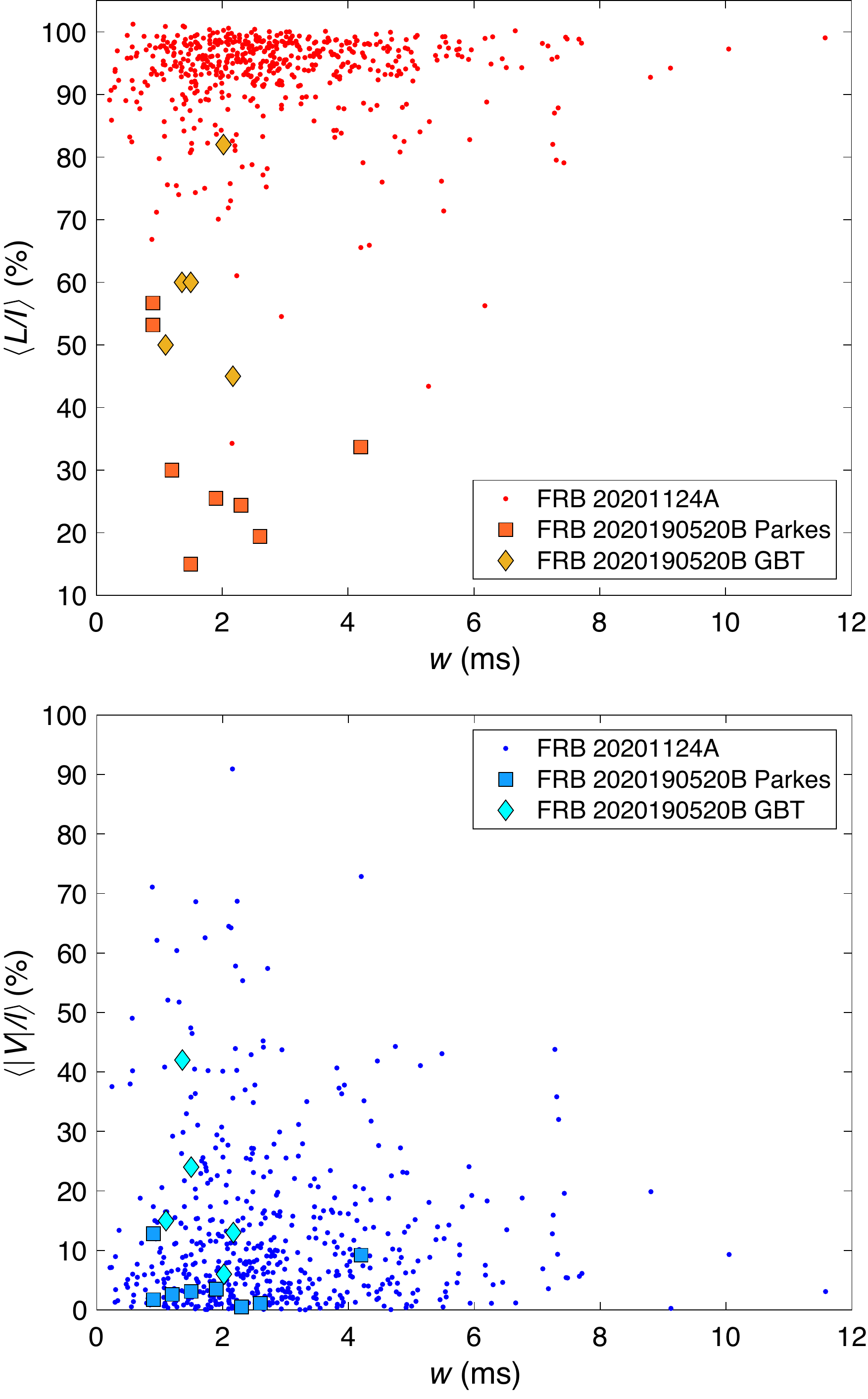}
\caption{LP/CP fraction as a function of pulse width. Data are quoted from \citet{Anna22,Dai22,Jiang22}.}
\label{fig7}
\end{figure}

FRB 20201124A is a highly active repeater. The source enters a newly active episode during September to October 2021.
A series of works on FRB 20201124A focused on the morphology \citep{Zhou22}, energy distribution \citep{Zhang22b}, polarisation \citep{Jiang22} and waiting time properties \citep{Niu22b} during the active episode.
Narrow emission spectra and a lot of drifting pattern events from the bursts are found, which are similar to other repeaters (e.g., \citealt{Pleunis21a}).
Bursts during the active episode exhibit an exponentially connected broken-power-law cumulative burst energy distribution, and a double-peak distribution of the waiting time, with no-detection of a credible spin period.
The source exhibits a variety of polarisation properties.
More than 90\% of the bursts have a total degree of polarisation larger than 90\%, and the average value among the sample is $(97.0 \pm 0.5)\%$, suggesting that most bulks may have an opening angle $\varphi_t>1/\gamma$.
The luminosities for most bursts are estimated as $\sim10^{39}\,\rm erg\,s^{-1}$ so that the waves can escape intact rather than scattered by electron-positron pairs in the magnetosphere \citep{Qu22}. 

Among 556 bright bursts, dozens show apparent sign change of CP.
However, these bursts also have sub-pulse structures, which exhibit drifting patterns (e.g., \citealt{Hessels19}).
The interval of the neighboring sub-pulses with different signs of CP is comparable with the sub-pulse duration, so that the CP sign change occurs at the bridge between sub-pulses.
The sign change of CP from sub-pulses are caused by multiple bulks of bunches. This is essentially different from the scenario discussed in Section \ref{sec2}.
These sub-pulses are produced by separate bulks of bunches during one trigger \citep{Wang20}.
The bursts of FRB 20201124A with high CP fractions do not show sign change of CP and the largest CP fraction can reach $\sim90\%$ \citep{Jiang22}. This is in tension with the ``thick'' bulk scenario.

We assume that the bulk length has $L=A\varphi_t$.
The duration of trigger is proportional to the size of the triggering region when $A$ is a constant.
For the ``thick'' bulk case, the average CP fraction is larger as the pulse width decrease.
However, for the ``thin'' bulk case, a burst (without sign change of CP) with certain pulse width can correspond to many $\left\langle |V|/I\right\rangle$, and the smallest value of them $\left\langle |V|/I\right\rangle_{\rm min}$ decreases with $\varphi_t$.
One can infer that the possibility of sign change of CP increases as $\varphi_t$ gets larger (for $\varphi_t\lesssim1/\gamma$).
We investigate the average LP and CP fractions as functions of pulse width for FRB 20201124A, as shown in Figure \ref{fig7}.
Although the sample with high CP is rare, there is no apparent trend that all of $\left\langle L/I\right\rangle$, $\left\langle |V|/I\right\rangle$ and $\left\langle |V|/I\right\rangle_{\rm min}$ evolve with pulse width, which is inconsistent with a constant $A$ for both of the ``thick'' and ``thin'' bulk cases.

\subsection{FRB 20121102A}

FRB 20121102A is the first confirmed repeater and has been intensively studied in  polarisation measurements.
The source has been measured to have a large and variable Faraday rotation measure, which is indicative of a complex magneto environment.
The bursts are $\sim$ 100\% linearly polarized with flat PA curves as measured at $4-8$ GHz \citep{Michilli18,Gajjar18}.
Depolarisation could be neglected according to the RM scattering relation extrapolated to L-band \citep{Feng22}.
The non-detection of CP for FRB 20121102A is consistent with the ``thin'' bulk case, which has a large opening angle ($\varphi_t\gg1/\gamma$, e.g., panel (b) of Figure \ref{fig4}).

The flat PA across burst profiles may be the consequence of a small horizontal-size bulk or a slow rotating neutron star.
A triple subpulse burst (burst 6, MJD 58075) shows dramatically variable PAs across burst profile \citep{Hilmarsson21}.
The burst has an upward-drifting pattern between the first two components, while downward drifting pattern for the second and third components.
This could be well understood if the LOS crosses the minimum impact angle.

\subsection{FRB 20180916B and 20190520B}

FRB 20190520B has been regularly detected as active during the several-month monitoring by FAST \citep{Niu22a}.
The total polarisation fraction of the source is smaller than $100\%$, indicating that significant absorption may occur at the complex magneto environment.
Most of the bursts are dominated by LP and a sign change of CP has not been reported.
These similar polarisation properties were also found in FRB 20180916 \citep{Nimmo21,Sand21,Pleunis21}.
All these observations are consistent with a bulk of $\varphi_t>1/\gamma$. However, it is hard to know whether the bulks are ``thick'' or ``thin''.
Similar with FRB 20201124A, we also investigate the average LP and CP fractions as functions of pulse width for FRB 20190520B at C-band, as shown in Figure \ref{fig7}.
All of $\left\langle L/I\right\rangle$, $\left\langle |V|/I\right\rangle$ and $\left\langle |V|/I\right\rangle_{\rm min}$ are independent of pulse width, which is not consistent with the duration of trigger being proportional to the size of the triggering region.
Both FRB 20180916B and FRB 20190520B have a total polarisation degree smaller than 100\% due to some possible propagation effects.
The general trend for both sources is that most bursts are dominated by LP, which may be caused by the intrinsic mechanism.

\section{Discussion and conclusions}\label{sec4}

We investigated the polarisation properties of coherent curvature radiation from charged bunches in the magnetosphere of a highly magnetized neutron star and applied the model
to interpret the polarized emission properties of repeating FRBs.
Emission from a charged bulk of bunches can exhibit a variety of polarisation properties.
We consider a more general expression of the angle of the emission cone by deriving the transverse bunch size.
If the LOS is confined to $\theta_b$, i.e. $\varphi<\theta_b$ (on-beam case), the observed burst retains high linear polarisation.
The bursts with high levels of CP tend to have low flux when $\varphi>\theta_b$ (off-beam cases) under the same condition.
As shown in Figure \ref{fig2}, CP-dominated waves appear at both sides of the beam, where the total flux can be up to the same magnitude as the peak of the profile.
However, if the fluctuation of any of $N_e$, $N_b$ or $\varphi_t$ exceeds an order of magnitude, the observed flux of waves would have more than one order of scattering, so that the trend of faint burst with high CP may be not apparent.

We apply this mechanism to explain the polarisation properties of repeating FRBs by considering two scenarios: (a) ``thick'' bulk ($t_{\rm int}<\theta_{\rm rad}/\Omega$); and (b) ``thin'' bulk ($\theta_{\rm rad}/\Omega<t_{\rm int}$).
The whole beaming radiation from the bulk could be observed for the ``thick'' bulk.
The average CP fraction of the burst is frequency dependent which can be up to tens of percent for $\varphi_t\lesssim1/\gamma$ while decrease to be smaller than $10\%$ when $\varphi_t\gg1/\gamma$.
For $\omega=\omega_c$, the average CP fraction is smaller than 55\% and it decreases as the pulse widths become larger.
However, one may only observe part of the radiation beaming region when it is an apparently ``thin'' bulk.
If so, highly circularly polarized emission ($\left\langle |V|/I\right\rangle>55\%$) might also be observed.
Bursts with CP sign change would be more circularly polarized as the observational window becomes wider.
Consequently, the model predicts that it is hard to observe a burst which has $\left\langle |V|/I\right\rangle>55\%$ with sign change of CP near the central frequency.
If the emission is generated from the same pole, the orientation of sign change of CP would also be the same.

Most FRBs are dominated by LP and rare events have sign change of CP.
The ``thin'' bulks with $\varphi_t>1/\gamma$ are the most likely cases for most bursts but the condition of $\varphi_t\lesssim1/\gamma$ may work for at least some bursts.
We investigated the average LP and CP fractions as functions of pulse width for FRB 20201124A and FRB 20190520B, and find no apparent trend for all of $\left\langle L/I\right\rangle$, $\left\langle |V|/I\right\rangle$ and $\left\langle |V|/I\right\rangle_{\rm min}$ in terms of the width.
The radial size of the emitting bulk does not seem to be proportional to its transverse size, so that the duration of an FRB trigger mechanism may not be directly related to the size of the triggering region.
The size of the FRB triggering region is comparable with the transverse bulk size on the stellar surface, i.e., $L_t\gtrsim R\Phi_t=3.1\times10^{3}R_6w_{-3}P_0^{-1}\sin\zeta/\sin\theta$ cm, where $R$ is the radius of the neutron star.
A high-tension point discharge on the surface may trigger bunches of electron-positron pairs during the oscillation-driven magnetospheric activity due to starquakes \citep{Lin15}.
The different bulk cases may be caused by different mechanisms to trigger energetic bunches on a neutron star surface.

For FRB 20201124A, the burst with $\left\langle |V|/I\right\rangle\approx90\%$ may be attributed to the observational window appearing at the side of the beam.
However, some bursts show polarisation profiles oscillating with wavelength due to radiative transfer \citep{Xu21}.
We cannot exclude the possibility that propagation effects contribute to CP for some bursts.

The intrinsic time depends on the number of bunches that travel through as the LOS sweeps the radiation region.
The bulk length may become larger as the bulk moves to a higher altitude, so that the continuous
plasma flow emits for a duration of $t_{\text{int}}\propto r\propto\nu^{-1}$.
Consequently, for sub-pulses emitted by the ``thin'' bulks with drifting structure, the drift rate is $\dot{\nu}\propto \nu/w\propto\nu^2$ \citep{Wang22}.
The relationship is different from a ``thick'' bulk emitter, in which $w=\theta_{\rm rad}/\Omega$, thus, $\dot{\nu}\propto\nu$.
Bursts are thought to be emitted at low heights, where multipolar fields may exist (e.g., \citealt{Bilous19,Kalapotharakos21}), leading to non-detection of periodicity of the underlying neutron star and a complex magnetic configuration.

Emission from charged bunches that is projected in the horizontal plane would be added coherently and the waves would have 100\% total polarisation degree.
However, fluctuations of photon arrival delay could arise from the charges moving in different trajectories due to the curved magnetic field lines, so that waves are added slightly incoherently.
The emission at the polarisation profile boundary is consequently depolarized ($\sqrt{L^2+V^2}/I<100\%$).
Alternatively, depolarization may also caused by finite temporal and spectral resolution or propagation effects \citep{Beniamini22}.

Since the plane of LP is determined by the orientation of the local magnetic field, PAs track down the geometry of the magnetic field lines in the emission region.
Most FRBs have flat PA curves but some others show varying PAs.
Within the magnetospheric model, the observed PA is calculated as $\Psi=\psi+\psi_{\rm RVM}$, where $\psi_{\rm RVM}$ is given by the rotation vector model \citep{Radhakrishnan69}.
The PA is generally flat for a slow-rotating pulsar, but
is varying when the impact angle is the smallest.
The bulk with a large $\Delta \chi_{ij}^2$ can also bring varying PAs across the burst profile.
The radial size of a bunch should be comparable or smaller than the half wavelength in order to allow coherent radiation, but the transverse size is at least $\gamma\lambda$.
The transverse size can be as large as the Fresnel length $\sim\sqrt{r\lambda}$, leading to $\Delta \chi\sim10^{-3}(\nu_9r_7)^{1/2}$.
Another radiation model invoked relativistic shocks far outside the magnetosphere can explain constant PA curves when the upstream ordered magnetic field has a fixed direction \citep{Metzger19}.
However, the model requires a fine-tuned magnetic field configuration to interpret the PA variation and is challenged by the observation of the high CP.

An FRB object may have a vacuum-like clean magnetosphere during a radio-quiet state, but electronic plasma can suddenly erupt from the stellar surface when the star becomes active.
The curvature model requires that positively (or negatively) charged dominated bulk can be produced via some trigger mechanisms.
The formation of charge bulk is essentially a question relevant to the nature of neutron star surface.

\section*{Acknowledgements}
We are grateful to Ze-Nan Liu, Rui Luo, Jia-Rui Niu, Shuang-Qiang Wang, Yuan-Pei Yang, Yong-Kun Zhang, Xiaoping Zheng, De-Jiang Zhou and an anonymous referee for helpful comments and discussions.
This work was supported by the National SKA Program of China (2020SKA0120100) and the National
Key R\&D Program of China (2017YFA0402602).
W.-Y.W. is supported by a Boya Fellowship and the fellowship of China Postdoctoral Science Foundation No. 2021M700247 and No. 2022T150018.
K.L. and R.X. are supported by the Strategic Priority Research Program of CAS (XDB23010200).

\section*{Data Availability}
No new data were generated or analysed in support of this research.







\bsp	
\label{lastpage}
\end{document}